\documentclass[11pt,a4paper]{article}
\usepackage[a4paper, width = 150mm, top = 30mm, bottom = 30mm]{geometry}
\usepackage{jheppub}
\usepackage{float}

\begin{document}

\title{\boldmath An Orbital Curvature for ISCOs}

\author[a]{Poulami Mukherjee,} 
\affiliation[a]{Department of Physics, School of Basic Sciences, Indian Institute of Technology Bhubaneswar, Bhubaneswar, Odisha, 752050, India}

\author[a,b]{Chandrasekhar Bhamidipati,}
\affiliation[b]{ Institute of Physics of the Czech Academy of Sciences \& CEICO\\ 
			Na Slovance 2, 18221 Prague, Czech Republic}

\author[c]{Sudipta Mukherji}
\affiliation[c]{Physics and Applied Mathematics Unit, Indian Statistical Institute, Kolkata, India}

\emailAdd{rm38@iitbbs.ac.in}
\emailAdd{chandrasekhar@iitbbs.ac.in}
\emailAdd{mukherji@iopb.res.in}

\abstract{
Building on the fact that circular orbits are the fixed points of  the phase portraits arising from the geodesic equations on a black hole spacetime, we develop a general orbital geometry, which captures the stability structure of orbits. The construction uses the orbital energy, angular momentum and angular velocity as state space variables, with response functions encoding how the circular orbits react to changes in the black hole background and in their angular motion. The associated scalar curvature provides a geometric diagnostic of certain distinguished points of the circular orbit family, such as the innermost stable circular orbit (ISCO), where it becomes singular. Across the ISCO, the relevant orbital response functions change sign as the system passes from the stable to the unstable circular orbit branch, and are found to vanish at marginal stability. We illustrate the framework for Schwarzschild black holes in asymptotically flat four dimensions, and five dimensional Schwarzschild AdS spacetimes. In the four dimensional case, the ISCO is associated with a degeneration of the energy response sector, while in the five dimensional AdS case it is detected through the angular response. The AdS example also exhibits a separate curvature singularity at a point which coincides with the minimum temperature of the black hole background.  These results show that the orbital geometry provides a promising unified framework for characterizing the stability of circular orbits and criticality of ISCOs.

}
\maketitle
\newpage
\section{Introduction}

Strong gravitational fields distort spacetime in nontrivial ways.  Close to a black hole, for instance, this distortion  becomes so strong that  stable circular timelike geodesics cease to exist below a critical  radius.  The circular orbit with this threshold radius is known as the innermost stable circular orbit or ISCO \cite{chandrasekhar1998mathematical,abramowicz2013foundations}. For a Schwarzschild black hole, ISCO occurs at exactly six times the mass of the black hole (in geometric units)~\cite{Kaplan:1949}. As the ISCO carries signatures of the background spacetime,  studying it has been essential for uncovering several key properties of the black holes \cite{Bardeen1972,PageThorne1974,Thorne1974,hartle2003,Bambi2017,Bambi2018}.\\

\noindent For a stationary spherically symmetric black hole in four or higher dimensions, equatorial geodesics of a massive test particle are parametrized  by its  energy $E$  and  angular momentum $L$. Tuning these control parameters yields either periodic orbits or trajectories that plunge directly into the black hole.  Within this dynamical system, circular orbits emerge as the fixed points of the governing geodesic equations \cite{dean1999variational,dean1999phase,bhamidipati2026criticality}. For large values of $L$, two distinct types of circular orbits exist: a stable orbit called a center, and an unstable saddle. As the angular momentum is reduced, these two fixed points approach one another and, finally,  at a critical threshold, they coalesce and disappear causing a saddle-node bifurcation~\cite{Strogatz:2015}. This is the point that precisely marks the ISCO.\\

 \noindent 
 It is well known that bifurcations in dynamical systems possess structural similarities with thermodynamic phase transitions. It is therefore natural to expect that  energy, angular momentum, angular velocity as well as various other functions associated with the fixed points in the vicinity of the ISCO exhibit power law
scaling relations akin to those found in second order phase transitions. Indeed in \cite{bhamidipati2026criticality}, such scaling relations 
were found and, subsequently, the critical exponents were extracted. The exponents were shown to acquire the same values as those arising from the mean field approximation of thermodynamic phase transitions.\\

\noindent 
An alternative approach to studying thermodynamic phase transitions is through the construction of a geometric space. By treating the equilibrium states of a system 
as points on a continuous Riemannian manifold, a metric can be defined using the second derivatives of an appropriate thermodynamic potential~\cite{Salamon:1983}.
Geometric quantities, particularly the scalar curvature associated with the manifold,  are  known to provide information about  correlations, and  instabilities of thermodynamic systems near the critical points. Owing to the analogy between bifurcations and phase transitions, it is  natural to inquire whether the instabilities at the ISCO 
can be captured through a similar geometric description. The present work is an  attempt to address this question.\\

\noindent 
To this end, we begin with a brief description of thermodynamic geometry that is relevant to the present context.  It is well known that  the thermodynamic state space of a system can be equipped with a Riemannian metric. The most commonly used is the Ruppeiner metric  which is defined  from the Hessian of the entropy \cite{Weinhold1975,ruppeiner1979thermodynamics,jan1,ruppeiner1995riemannian},
\begin{eqnarray}
    g_{ab}= -\frac{\partial^2 S} {\partial X^a\partial X^b},
\end{eqnarray}
where $X^a$ denote the extensive variables of the system. With an appropriate choice of the thermodynamic variables, the components of the Ruppeiner metric can be expressed in terms of the familiar response functions, such as heat capacities and compressibilities, providing information on phase transitions ranging from standard thermodynamic systems to black holes~\cite{Cai:1998ep,Dolan:2002wm,Aman1,Janke:2004ef,Quevedo:2006xk,Ruppeiner:2010dzw,Sahay:2010tx,Hendi:2015rja,Mansoori:2016jer,Bhattacharya:2017hfj,Wei:2015iwa,Wei:2019uqg,Mansoori2025}.  The singularities or zeroes of the response functions are generally reflected in the scalar curvature of the thermodynamic state space, though, there can be exceptions~\cite{Davies:1977,Aman1,Aman:2005xk}, including ensemble dependent behavior~\cite{Bravetti:2012dn,Ghosh:2023khd,Aman:2005xk}. \\

\noindent 
In the present scenario, we note that the circular orbits are the fixed (or the equilibrium) points of  the phase portraits arising from the geodesic equations on a black hole spacetime. The Hawking temperature ($T_H$) of the black hole then naturally gets associated with these orbits. The other relevant quantities are the energy $E$, the angular momentum $L$ and the angular velocity  $\Omega$ of the orbiting particle. By treating the circular orbits, i.e., the fixed points of the radial dynamics, as points of a two dimensional manifold, we can define a metric on the orbital state space. Unlike the Ruppeiner metric, this metric is not positive definite in general, and its signature changes across the branches of circular orbits. We find two important response functions sensitive to the changes in the background as well as in the orbital parameters. These are 
\begin{equation}
C_\Omega
=
\left(\frac{\partial E}{\partial T_H}\right)_\Omega,
\label{eq:COmega_general}
\end{equation}
and
\begin{equation}
\chi_E= -\left(\frac{\partial L}{\partial\Omega}\right)_E.
\label{eq:chiE_general}
\end{equation}
While $C_\Omega$ measures the response of the orbital energy $E$ to the Hawking temperature $T_H$ at a fixed angular velocity,  $\chi_E$
measures the angular response at a fixed energy. We will see later that the vanishing of these functions is closely correlated with the marginal 
stability of the circular orbits. At this point, to avoid any future ambiguities, we emphasize that the state space considered here is {\it not} the thermodynamic state space of the black hole. Instead, it is the space of all the circular timelike orbits, with $(E, \Omega)$ used as coordinates. We refer to the resulting construction as the orbital geometry. \\

\noindent 
In the following, we consider circular geodesics of a massive test particle  on a one parameter family of static and spherically symmetric geometries written in Schwarzschild type coordinates,
\begin{equation}
    ds^2
    =-f(r;p)\,dt^2 + \frac{dr^2}{f(r;p)} + r^2d\Sigma_{d-2}^2,
\label{eq:metric}
\end{equation}
where $d\Sigma_{d-2}^2$ is the line element of the unit $(d-2)$-sphere and $p$ denotes a parameter of the black hole family. The general construction of an explicit line element of the orbital geometry and the precise definitions entering into its construction are presented in section~\ref{sec:orbital_geometry}. This is done  without committing to a particular form of $f(r;p)$. The resulting construction raises a simple question: can the curvature of the orbital metric capture the loss of stability at the ISCO? The answer is not determined by the behavior of any one response function alone. The curvature depends on both metric components and on their derivatives, so a divergence in $C_\Omega$ need not produce a curvature singularity, while a zero of a response function can make the metric degenerate, and lead to a genuine curvature divergence. Section~\ref{sec:orbital_geometry} ends with an elaboration of these issues. Next, we specialize to four dimensional asymptotically flat Schwarzschild and five dimensional  AdS  Schwarzschild black holes.  Analysis for these cases is provided in section~\ref{sec:schwarzschild} and section~\ref{sec:schwarzschild_ads}, respectively. In four dimensional Schwarzschild spacetime~\cite{chandrasekhar1998mathematical,song2021evolutions}, orbital curvature captures the ISCO as the critical point of the circular orbit family, whereas in 5D Schwarzschild AdS spacetime \cite{gonzalez2023time}, this also retains information about the temperature structure of the black hole apart from the ISCO. The resulting curvature consequently provides more than a restatement of the ISCO condition. It records how different orbital response functions become singular, vanish, or remain finite at physically distinguished points. We use this behavior to identify which curvature features are tied directly to marginal stability and which arise from the thermodynamic response of the background. In this sense, the orbital curvature gives a geometric characterization of circular orbit response.
We conclude in section~\ref{sec:conclusion} by summarizing the main results and discussing the possible use of orbital response geometry for other black hole backgrounds, as well as some open questions. The curvature calculation and other relations used in the analysis of this paper are collected in appendix~\ref{app:curvature}.

\section{General construction of the orbital geometry}
\label{sec:orbital_geometry}

Starting with a general black hole metric  of the form (\ref{eq:metric}), in this section, we first construct the line element pertaining  to the  
orbital geometry. We do not assume any specific form of the lapse function $f(r;p)$. Our aim is to extract some general features
of the orbital geometry as we approach the point of marginal orbital stability.\\

\noindent Given the form of the metric, the outer horizon 
$r=r_h(p)$  is determined by the largest  positive root of the equation
\begin{equation}
f(r_h;p)=0.
\end{equation}
The corresponding Hawking temperature is
\begin{equation}
T_H=\frac{f_r(r_h;p)}{4\pi},
\label{eq:TH_general}
\end{equation}
where a subscript $r$ denotes differentiation with respect to $r$.
In a timelike circular orbit at a radius $r=r_c$, the energy, angular momentum and angular velocity can be expressed directly in terms of the lapse function and its derivatives evaluated at $r_c$,
\begin{equation}
E^2=\frac{2f^2}{2f-r_cf_r},
\qquad
L^2=\frac{r_c^3f_r}{2f-r_cf_r},
\qquad
\Omega^2=\frac{f_r}{2r_c}.
\label{eq:orbit_quantities}
\end{equation}
We use units with $G=c=\hbar=k_B=1$, and $E$ and $L$ denote the energy and angular momentum per unit rest mass of the particle. These quantities provide a convenient parameterization of the circular orbit
state space. The stability of a circular orbit is determined by the effective potential~\cite{Cardoso:2008bp}. Equivalently, we can examine the variation of the angular momentum along the circular orbits as,
\begin{equation}
\frac{\partial L^2}{\partial r_c}=\frac{2r_c^2 K}{(2f-r_cf_r)^2},
\quad \text{where} \quad K= 3ff_r+r_c f f_{rr}-2r_cf_r^2\, .
\label{eq:dL2drc}
\end{equation}
Thus, the marginal stability condition is 
\begin{equation}
K=3ff_r+r_c f f_{rr}-2r_cf_r^2=0.
\label{eq:ISCO_general}
\end{equation}
This condition identifies the ISCO. On the two sides of the ISCO, the circular orbits are separated into the stable ($K>0$) and unstable ($K<0$) branches. The corresponding fixed points of the orbital dynamics are a center and a saddle, respectively. At the ISCO these two branches meet and the fixed points coalesce. According to \cite{bhamidipati2026criticality} the critical behavior of the orbital quantities here has a close analogy with a van der Waals system. 
This correspondence suggests that the ISCO critical point may also admit a geometric description in terms of suitable orbital response functions, which were alluded to earlier in eqns. (\ref{eq:COmega_general}) and (\ref{eq:chiE_general}).   \\

\noindent
Motivated by developments in thermodynamic geometry~\cite{Cai:1998ep,Dolan:2002wm,Aman1,Janke:2004ef,Quevedo:2006xk,Ruppeiner:2010dzw,Sahay:2010tx,Bravetti:2012dn,Hendi:2015rja,Mansoori:2016jer,Bhattacharya:2017hfj,Wei:2019uqg,Ghosh:2023khd,Mansoori2025}, we propose a two dimensional metric on the orbital state space parameterized by $(E,\Omega)$,
\begin{equation}
ds_{\rm orb}^2= F(E,\Omega)\,dE^2+G(E,\Omega)\,d\Omega^2,
\label{eq:orbital_metric}
\end{equation}
with
\begin{equation}
F=\frac{\alpha\,C_\Omega}{E^2},
\qquad
G=\frac{\chi_E}{E}.
\label{eq:FG_definition}
\end{equation}
Here $\alpha>0$ is a quantity with dimensions of inverse length, introduced so that both terms of the line element are dimensionless, and will be chosen appropriately in the sequel. 
The scalar curvature of this two dimensional geometry can be evaluated without specifying a particular black hole solution. For a two dimensional diagonal metric of the form in eqn.~\eqref{eq:orbital_metric}, the curvature takes the form (see appendix~\ref{app:curvature} for the detailed calculation), 
\begin{equation}
R= -\frac{1}{FG} \left[G_{,EE}+F_{,\Omega\Omega} -\frac{1}{2}G_{,E}\partial_E\ln|FG|-\frac{1}{2}F_{,\Omega}\partial_\Omega\ln|FG|\right].
\label{eq:curvature_general}
\end{equation}
This is the form used in the rest of the paper. A useful combination in this expression is
\begin{equation}
FG= \det g= \frac{\alpha\,C_\Omega\chi_E}{E^3}.
\label{eq:FG_product}
\end{equation}
Some information about the possible singular behavior of the orbital curvature already ensues from eqn.~\eqref{eq:FG_product}, without actually committing to any specific background, which we record below:
\begin{itemize}
\item $C_\Omega=0$ implies $F=0$ and is a candidate curvature singularity. We will see later that this is what happens at the ISCO, when we consider asymptotically flat Schwarzschild black holes in four dimensions.
\item $\chi_E=0$ implies $G=0$ and gives an angular sector degeneration. As we discuss later, this is the mechanism by which the orbital curvature diverges at the ISCO point in the Schwarzschild black holes in AdS$_5$.
\item Divergence of $C_\Omega$ does not by itself imply a divergence of $R$, since the prefactor $1/(FG)$ and the derivative terms in eqn.~\eqref{eq:curvature_general} must be considered together to decide whether the curvature actually diverges or not.
\item Another point worth noting from eqn. (\ref{eq:orbit_quantities}) is that the denominator of the energy and angular momentum, $2f-r_cf_r=0$ is the photon sphere boundary, where $E$ and $L$ diverge. Thus, one expects the curvature $R$ to also show special behavior in these limits. This will be discussed on a case-by-case basis.

\item In general, for the calculation of curvature, it will often be useful to transform the coordinates. The map may be singular at the ISCO point, which leads to possible divergences of $R$. This will be addressed separately.
\end{itemize}
The two examples to follow realize these mechanisms differently.

\section{Four dimensional Schwarzschild black holes}
\label{sec:schwarzschild}

\begin{figure}[H]
    \centering
    \includegraphics[width=0.75\textwidth]{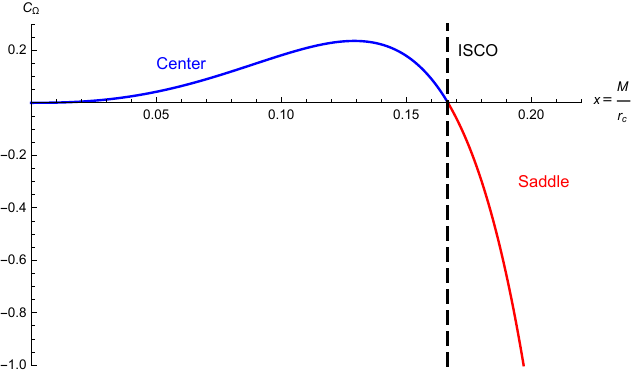}
    \caption{Orbital response $C_\Omega$ as a function of $x= M/r_c$ for $\Omega=0.1$ (other values of $\Omega$ only rescale the vertical axis by a factor $0.1/\Omega$, since $C_\Omega\propto 1/\Omega$). The dashed vertical line is the ISCO, $x= 1/6$. The response function is positive on the center branch, vanishes at the ISCO, and is negative on the saddle branch. Since $\alpha=T_H>0$, the combination $T_HC_\Omega$ entering the orbital metric has the same sign structure.}
    \label{fig:Comega_4D}
\end{figure}

The first case we consider is four dimensional asymptotically flat Schwarzschild geometry~\cite{chandrasekhar1998mathematical,song2021evolutions},
\begin{equation}
    f(r)=1-\frac{2M}{r},
    \qquad
    r_h=2M,
    \qquad
    T_H=\frac{1}{8\pi M}.
\end{equation}
Here we apply the orbital geometry introduced in section~\ref{sec:orbital_geometry} to the four dimensional asymptotically flat Schwarzschild black hole to determine explicitly how the orbital response functions enter the geometry and, specifically how the orbital curvature behaves at the ISCO in this case. The circular orbit structure of this geometry is well known, and the corresponding phase plane description exhibits a center and a saddle which
coalesce at the ISCO, at $ r_{\rm ISCO}=6M$ \cite{dean1999phase,jia2018existence,bhamidipati2026criticality}. We use the orbit variable defined in \cite{bhamidipati2026criticality}:
\begin{equation}
    x = \frac{M}{r_c}.
\end{equation}
In terms of $x$, the timelike circular orbit region is
\begin{equation}
    0<x<\frac{1}{3}.
\end{equation}
The two portions of this interval are separated by the ISCO,
\begin{equation}
    0<x<\frac{1}{6} \quad \text{stable}, 
    \qquad
    x=\frac{1}{6} \quad \text{ISCO},
    \qquad
    \frac{1}{6}<x<\frac{1}{3} \quad \text{unstable}.
\end{equation}
This follows from eqn.~\eqref{eq:dL2drc}, which here gives $K=2\Omega\sqrt{x}\,(1-6x)$.
The endpoint $x=1/3$ corresponds to the photon orbit boundary, which lies outside the timelike region, and where the timelike circular orbit description ceases to apply.
In what follows, the stable and unstable branches will be kept distinct, with the ISCO treated as a limiting point of the two open branches. This is also the standard center saddle structure of the circular orbit dynamics discussed in phase space treatments of ISCOs \cite{jia2018existence,bhamidipati2026criticality}. The energy, angular momentum, and angular velocity of a circular timelike orbit follow from the general expressions in section~\ref{sec:orbital_geometry}. In terms of the variables $(x,\Omega)$ they take the form
\begin{equation}
    E(x)=\frac{1-2x}{\sqrt{1-3x}}, \qquad
    L(x,\Omega)
    =\frac{x}{\Omega\sqrt{1-3x}},
\qquad
    \Omega^2
    =\frac{M}{r_c^3}
    =\frac{x^3}{M^2}.
\end{equation}
It is therefore possible to use $(x,\Omega)$ in place of $(r_c,M)$, as
\begin{equation}
    r_c=\frac{\sqrt{x}}{\Omega},
    \qquad
    M=\frac{x^{3/2}}{\Omega},
    \qquad
    r_h=\frac{2x^{3/2}}{\Omega}.
\end{equation}
We now evaluate the response functions entering the orbital metric. At fixed
angular velocity, the energy response to the Hawking temperature is
\begin{equation}
    C_\Omega =\left(\frac{\partial E}{\partial T_H}\right)_\Omega = \frac{4\pi x^2K}{\big(3\Omega^2(1-3x)^{3/2}\big)}\, ,
    \label{eq:sch_Comega}
\end{equation}
which implies that its sign is that of $K$. We now take $\alpha=T_H$, so that $\alpha C_\Omega=T_H C_\Omega=(\partial E/\partial \ln T_H)_\Omega$ is the logarithmic form of the energy response. This choice is natural for the current Schwarzschild geometry under study, which has no scale other than $M$\footnote{Notice that one could have equivalently made the choice $\alpha=1/M$, which differs only by a constant factor, as $T_H \frac{\partial}{\partial T_H} = - M \frac{\partial}{\partial M}$.}. With $\alpha=T_H$, the orbital metric is invariant under the rescaling $\Omega\to\lambda\Omega$ at fixed $x$ (equivalently $M\to M/\lambda$), so that the orbital geometry inherits the scaling symmetry of the background, and its curvature depends on $x$ alone, as we find below. In fact, the quantity entering the orbital metric~\eqref{eq:FG_definition} is
\begin{equation}
    T_H\,C_\Omega =\frac{x(1-6x)}{3(1-3x)^{3/2}},
    \label{eq:sch_ThComega}
\end{equation}
which is independent of $\Omega$. The zero of $T_HC_\Omega$ at the ISCO, which underlies the curvature divergence found below, follows directly from the factor $K$ in $C_\Omega$, in eqn. (\ref{eq:sch_Comega}). The factor $(1-6x)$ is important for the geometry. $C_\Omega$ vanishes precisely at
the ISCO and changes sign when the orbit changes from the stable to the
unstable branch. Thus, within the present construction, the response
function already distinguishes the two sides of the marginally stable
orbit, i.e.,
\begin{equation}
    C_\Omega>0
    \quad (x<1/6),
    \qquad
    C_\Omega=0
    \quad (x=1/6),
    \qquad
    C_\Omega<0
    \quad (x>1/6).
\end{equation}
Notice that $C_\Omega$ is an orbital response function and should not be
identified with the thermodynamic heat capacity of the black hole. The corresponding behavior of the response function is shown in figure~\ref{fig:Comega_4D}.
\noindent

\noindent
Using eqn.~\eqref{eq:sch_ThComega} in the orbital line element, we get
\begin{equation}
    ds_{\rm orb}^2=g_{EE}\,dE^2+g_{\Omega\Omega}\,d\Omega^2\, .
\end{equation}
The first component of the orbital metric is
\begin{equation}
    g_{EE} =\frac{T_H\,C_\Omega}{E^2} = \frac{x(1-6x)}{3(1-2x)^2\sqrt{1-3x}},
    \label{eq:sch_gEE}
\end{equation}
and the second component is obtained from the second response function as,
\begin{equation}
    g_{\Omega\Omega} =\frac{\chi_E}{E}=\frac{x}{\Omega^2(1-2x)}.
    \label{eq:sch_gOmegaOmega}
\end{equation}
The two components therefore respond differently to the ISCO. The metric component $g_{EE}$ vanishes linearly at $x=1/6$, whereas the angular component remains finite there. This degeneracy is the feature that controls the curvature near the marginally stable orbit. The curvature associated with this metric will show whether the zero of the response develops into a genuine geometric singularity or not.\\

\noindent
For the Schwarzschild solution, $E$ depends only on $x$ and
\begin{equation}
    \frac{dE}{dx}=\frac{6x-1}{2(1-3x)^{3/2}}.
    \label{eq:sch_dEdx}
\end{equation}
It is therefore convenient to pull back the orbital metric to the coordinates $(x,\Omega)$. The corresponding Jacobian is
\begin{equation}
    \mathcal{J}= \frac{\partial(E,\Omega)}{\partial(x,\Omega)} 
    =\begin{pmatrix}
    dE/dx & 0\\0 & 1
    \end{pmatrix},
\end{equation}
and the pullback of the metric becomes
\begin{equation}
    d\tilde s_{\rm orb}^2=\tilde g_{xx}\,dx^2 +\tilde g_{\Omega\Omega}\,d\Omega^2,
\end{equation}
where
\begin{equation}
    \tilde g_{xx} =\frac{x(1-6x)^3}{12(1-2x)^2(1-3x)^{7/2}},
    \label{eq:sch_gxx}
\end{equation}
\begin{equation}
    \tilde g_{x\Omega}=0,
    \qquad
    \tilde g_{\Omega\Omega}= \frac{x}{\Omega^2(1-2x)}.
    \label{eq:sch_gxomega}
\end{equation}
The pullback is considered separately on the stable and unstable open branches, with the ISCO approached as a limiting point. The coordinate transformation becomes degenerate at the ISCO because $dE/dx$ vanishes there, and the curvature at the ISCO is consequently obtained by approaching the critical point from either branch.\\

\noindent
Now, substituting the response functions into the two dimensional curvature
formula  in eqn. (\ref{eq:curvature_general}) (see appendix~\ref{app:curvature-4d} for details) gives the following expression:
\begin{equation}
    R(x)=-\frac{3(1-3x)^{5/2}\left(180x^3-192x^2+63x-4\right)}{x^3(6x-1)^4}.
\label{eq:sch_R}
\end{equation}
Eqn.~\eqref{eq:sch_R} shows that the orbital curvature is sensitive to the marginal stability of the circular orbit. The singular part is controlled by $(6x-1)^{-4}$, while the
remaining factors stay finite at $x=1/6$. Behavior of the orbital curvature is shown in figure~\ref{fig:R_4D}, and some of its significant features are as follows:
\begin{figure}[H]
    \centering
    \includegraphics[width=0.75\textwidth]{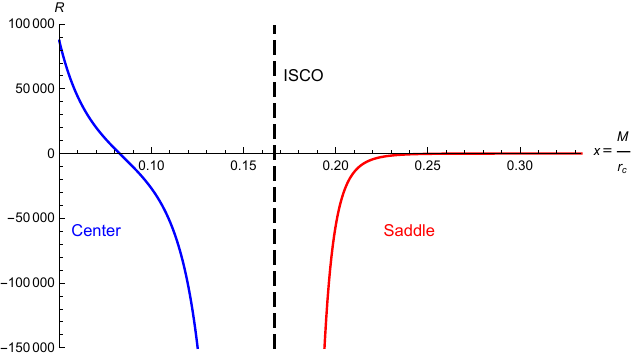}
    \caption{Ricci scalar $R$ as a function of $x= M/r_c$. The dashed vertical line marks the ISCO at $x= 1/6$, while the center (stable) branch is in blue, and the saddle (unstable) branch in red.}
    \label{fig:R_4D}
\end{figure}
\begin{itemize}
    \item Near the ISCO, the response function and the energy derivative behave as
\begin{equation}
    T_HC_\Omega \simeq -\frac{2\sqrt{2}}{3}\left(x-\frac16\right),\qquad
    \frac{dE}{dx}\simeq 6\sqrt{2}\left(x-\frac16\right)\, ,
    \label{Emin}
\end{equation}
so that $E(x)$ has a minimum at the ISCO, with $E-E_{\rm ISCO}\simeq 3\sqrt{2}\,(x-1/6)^2$ and $E_{\rm ISCO}=\sqrt{8/9}$ (see eqn. (2.24) of~\cite{bhamidipati2026criticality})). It then follows from eqn.~\eqref{eq:sch_R} that
\begin{equation}
    R\simeq-\frac{\sqrt{2}}{8}\left(x-\frac16\right)^{-4}
    \simeq-\frac{9\sqrt{2}}{4}\,\frac{1}{\left(E-E_{\rm ISCO}\right)^{2}}\,,
    \qquad x\rightarrow\frac16\,.
    \label{eq:sch_R_EISCO}
\end{equation}
Hence $R\to-\infty$ on both branches, as $x\to\frac16^{\pm}$, or equivalently as $E\to E_{\rm ISCO}$ from above. The exponent $4$ in $x$ becomes $2$ in the energy, because $E(x)$ is quadratic about its minimum. The curvature therefore diverges from both sides of the ISCO, whereas $C_\Omega$ changes sign across the same point. The curvature singularity thus occurs precisely where the stable and unstable circular orbit branches meet. In this sense the curvature provides a unified geometric treatment of the marginal stability condition, and stability features of the circular orbit response.
\item It is also useful to examine the behavior away from the ISCO. As the photon orbit boundary is approached,
\begin{equation}
    x\rightarrow\frac13^{-},
\end{equation}
the response grows without bound, while the curvature vanishes, i.e., 
\begin{equation}
    T_HC_\Omega\rightarrow-\infty,
    \qquad
    R\simeq-189\,(1-3x)^{5/2}\rightarrow 0.
\label{eq:sch_photon}
\end{equation}
Thus the divergence of the response function does not by itself imply a curvature singularity. The factors entering the curvature combine in such a way that the leading divergence cancels as the photon orbit is approached. This behavior is distinct from the ISCO, where the curvature itself develops a genuine divergence.
\item At the opposite end of the circular orbit interval,
\begin{equation}
    x\rightarrow0,
\end{equation}
one finds
\begin{equation}
    R
    \sim
    \frac{12}{x^{3}},
    \qquad
    T_HC_\Omega
    \sim
    \frac{x}{3}.
\label{eq:sch_infinity}
\end{equation}
This limit corresponds to $r_c\rightarrow\infty$ rather than to a finite radius critical orbit. The resulting curvature divergence should therefore be regarded as an asymptotic endpoint of the chosen orbital coordinates, rather than as an additional orbital transition.
\item Between these two limits, the curvature changes sign exactly once. From eqn.~\eqref{eq:sch_R}, $R$ vanishes in the range $0<x<1/3$ at $x_0\simeq 0.0827$ (i.e., $r_c\simeq 12.09\,M$),
which lies on the stable branch. Thus $R>0$ for $x<x_0$, while $R<0$ for $x_0<x<1/3$, including the whole unstable branch. Unlike the ISCO, this zero is not accompanied by any special behavior of the response functions, and does not presently correspond to a known radius for the Schwarzschild black holes.\\
\end{itemize}
\noindent
For comparison, it is useful to note that the usual  thermodynamic heat capacity of the Schwarzschild black hole is
\begin{equation}
    C_{\rm BH} =\frac{dM}{dT_H}=-8\pi M^2.
\end{equation}
This quantity is conceptually different from $C_\Omega$. The latter describes the response of the energy of a circular orbit to the background Hawking temperature at fixed orbital angular velocity. The distinction is important because the curvature singularity found above is a property of the proposed orbital state space geometry and should not be interpreted as the usual thermodynamic instability of the Schwarzschild black hole.\\

\noindent
This example therefore provides a simple realization of the orbital geometry proposed in this work. The ISCO is characterized simultaneously by the vanishing of the orbital response and by a divergence of the associated Ricci scalar. In contrast, the divergence of $C_\Omega$ as the photon orbit is approached does not lead to a curvature singularity. These two behaviors will be useful for comparison with the five dimensional Schwarzschild AdS geometry examined in the next section.

\section{Five dimensional Schwarzschild AdS spacetime}
\label{sec:schwarzschild_ads}

We now consider the five dimensional Schwarzschild AdS geometry. The classification of circular orbits as center and saddle, together with the ISCO condition, and also thermodynamics of Schwarzschild AdS black holes is well known, beginning with the analysis of
the Hawking temperature and its minimum as a function of the horizon radius~\cite{hawking1983thermodynamics,Witten:1998zw,Cruz:2004ts,Hackmann:2008zza,gonzalez2023time,bhamidipati2026criticality}. Here we are interested in how these
features appear in the orbital geometry introduced in section~\ref{sec:orbital_geometry}.
The metric function is
\begin{equation}
f(r) = 1+\frac{r^2}{\ell^2} -\frac{M}{r^2},
\end{equation}
where from the condition for existence of the horizon, we have
\begin{equation}
    M =r_h^2 \left(1+\frac{r_h^2}{\ell^2}\right).
\end{equation}
The Hawking temperature is
\begin{equation}
    T_H =\frac{1}{2\pi r_h}\left(1+\frac{2r_h^2}{\ell^2}\right).
\end{equation}
The Hawking temperature is therefore a non-monotonic function of $r_h$. This feature becomes
important below since the orbital response $C_\Omega$ involves a derivative
with respect to $T_H$.\\

\begin{figure}[H]
\centering
\includegraphics[width=0.75\textwidth]{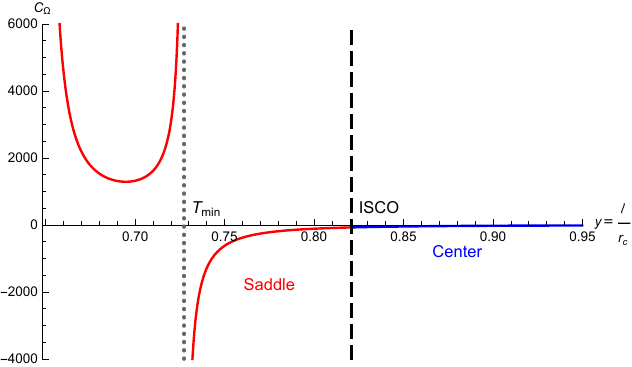}
    \caption{\footnotesize Orbital response $C_\Omega$ (in units of $\ell$) for the five dimensional Schwarzschild AdS case at fixed $w=\ell\Omega=1.1$, plotted against $y=\ell/r_c$. The red and blue curves denote the saddle and center branches, respectively. $C_\Omega$ diverges at the minimum temperature locus, $2h^2-1=0$ (gray dotted line), where it changes sign from positive (large black holes, $h>1/\sqrt2$) to negative (small black holes, $h<1/\sqrt2$). It also diverges as the timelike boundary $B=0$ is approached, and remains finite at the ISCO, $K=0$ (black dashed line).}
    \label{fig:Comega_ads}
\end{figure}

\noindent
It is convenient to introduce the dimensionless variables
\begin{equation}
    y=\frac{\ell}{r_c},
    \qquad
    h=\frac{r_h}{\ell},
    \qquad
    w=\ell\Omega,
    \qquad
    \lambda=\frac{L}{\ell},
    \qquad
    \tau=\ell T_H.
\end{equation}
For a circular orbit,
\begin{equation}
 \Omega^2=\frac{f'(r_c)}{2r_c}
 =\frac{1}{\ell^2}+\frac{M}{r_c^4}.
\end{equation}
One therefore obtains the condition
\begin{equation} \label{constraint}
 h^2(1+h^2)=\frac{w^2-1}{y^4}\, .
\end{equation}
The above equation is quite useful because it allows the horizon variable $h$ and the Hawking temperature to be expressed in terms of the orbital variables $(y,w)$,  as
\begin{equation}
  h^2=\frac{-1+\sqrt{1+4\frac{w^2-1}{y^4}}}{2}\, , \qquad  \tau =\frac{1+2h^2}{2\pi h}\, .
\end{equation}
As we will see below, this makes it possible to evaluate the response functions
directly in the same variables used to describe the circular orbits.  The condition in eqn. (\ref{constraint}) therefore provides the link between the background thermodynamics and the orbital geometry. In addition, for fixed $w$, the horizon radius can be determined by the orbital radius. Thus, 
the circular orbit quantities can be written succinctly as
\begin{equation}
    E=\frac{B+w^2}{y\sqrt{B}},
    \qquad
    \lambda=\frac{w}{y\sqrt{B}},
    \label{eq:orbit quantitites_5D}
\end{equation}
where we defined
\begin{equation}
     B=y^2+2-2w^2.
\end{equation}
The timelike circular orbit region is characterized by $B>0$ and will be imposed throughout the following analysis. In what follows, $(y,w)$ serve as coordinates on the two-dimensional orbital state space, related to $(E,w)$ through eqn.~(\ref{eq:orbit quantitites_5D}); the figures are shown at fixed $w$. For fixed $\ell$, fixing $w$ is equivalent to fixing the angular velocity
$\Omega$. The orbital response introduced in section~\ref{sec:orbital_geometry} can therefore be written as
\begin{equation}
    C_\Omega =\ell C_w,
    \qquad
    C_w = \left(\frac{\partial E}{\partial\tau}\right)_w =\frac{2\pi h^3 (1+2h^2)J y^3}{(w^2-1)(2h^2-1)B^{3/2}}\, ,
    \label{eq:Cw_ads}
\end{equation}
where
\begin{equation}
    J=B + w^2 (w^2-1)\, ,
    \end{equation}
    is always positive, since $B>0$ and $w^2>1$.
The denominator of $C_w$ contains a factor which is directly related to the derivative of
temperature, i.e.,
\begin{equation} 
    2h^2-1=0
    \qquad\Longleftrightarrow\qquad
    \frac{dT_H}{dr_h}=0.
    \label{tmin1}
\end{equation}
This is the minimum temperature locus of the Schwarzschild AdS black hole. As a result, $C_w$ (and hence $C_\Omega$) diverges there. There is another  possible divergence as the timelike boundary $B=0$ is approached. For the present case of Schwarzschild AdS black hole, due to the availability of a natural AdS length scale $\ell$, we take $\alpha=1/\ell$. Then, the quantity entering the orbital metric is $\alpha C_\Omega=C_w$. The behavior of $C_\Omega=\ell C_w$, in units of $\ell$, is plotted in figure~\ref{fig:Comega_ads}.  \\

\begin{figure}[H]
\centering
\includegraphics[width=0.75\textwidth]{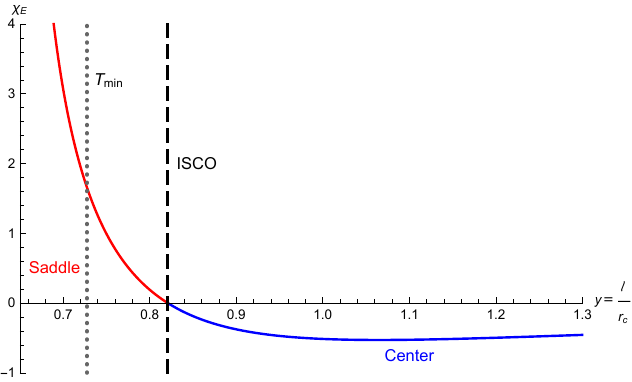}
    \caption{\footnotesize Fixed energy angular susceptibility $\chi_E$ for the five dimensional Schwarzschild AdS case at fixed $w=\ell\Omega=1.1$, plotted against $y=\ell/r_c$. The red and blue curves denote the saddle and center branches, respectively. $\chi_E$ is positive on the saddle branch, vanishes at the ISCO, $K=0$ (black dashed line), and is negative on the center branch. It remains finite at the minimum temperature locus, $2h^2-1=0$ (gray dotted line), and diverges as the timelike boundary $B=0$ is approached.}
    \label{fig:chiE_ads}
\end{figure}

\noindent
The second response function is
\begin{equation}
    \chi_E =-\left(\frac{\partial\lambda}{\partial w}\right)_E = -\frac{y^2 K}{8 J\sqrt{B}},
    \label{ChiE_5D}
\end{equation}
which is dimensionless, and is related to $-(\partial L/\partial\Omega)_E$ by a factor of $\ell^2$. For simplicity, we denote this dimensionless quantity also by $\chi_E$. Here $K$ is the marginal stability factor of eqn.~(\ref{eq:dL2drc}), evaluated in units of the AdS radius (i.e., with $r_c$ measured in units of $\ell$), which for the present background reads
\begin{equation}
K = \frac{8}{y^3}\left(2+y^2-w^2-w^4\right).
    \label{K_5D}
\end{equation}
Thus,
\begin{equation}
    K>0:\ {\rm stable},
    \qquad
    K=0:\ {\rm ISCO},
    \qquad
    K<0:\ {\rm unstable}.
\end{equation}
The ISCO is located at
\begin{equation}
    y_{\rm ISCO}^2
    =w^4+w^2-2,
    \qquad
    w>1.
    \label{eq:yISCO}
\end{equation}
Hence $\chi_E$ vanishes at the ISCO and changes sign across it, while $C_w$ remains finite there. This is different from the four dimensional Schwarzschild case, where the zero of the orbital temperature response occurs at the ISCO. The corresponding behavior of  $\chi_E$ is shown in figure~\ref{fig:chiE_ads}. \\

\begin{figure}[H]
    \centering
    \includegraphics[width=0.75\textwidth]{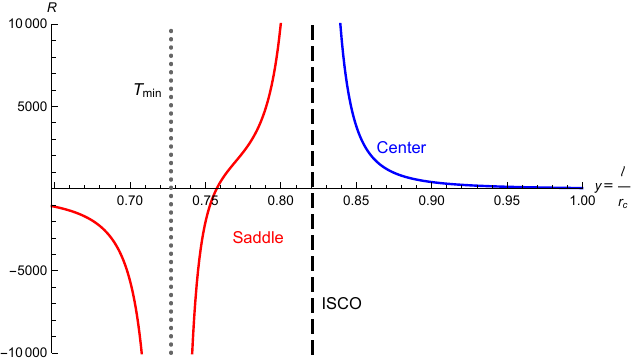}
    \caption{\footnotesize Scalar curvature $R$ for the five dimensional Schwarzschild AdS case for $w=1.1$. The black dashed line marks the ISCO, $K=0$, and the gray dotted line marks the minimum temperature locus, $2h^2-1=0$. The red and blue curves are the saddle and center sectors, respectively.}
    \label{fig:R_5D}
\end{figure}

\noindent
Since $\alpha>0$, $\alpha C_\Omega$ has the same sign, zeros and divergences as $C_\Omega$. The orbital metric of eqn. (\ref{eq:orbital_metric}) can now be written entirely in terms of the dimensionless variables, since $\alpha\,C_\Omega\, dE^2=C_w\,dE^2$ and $-(\partial L/\partial\Omega)_E\,d\Omega^2=\chi_E\, dw^2$. Using
\begin{equation}
    F=\frac{C_w}{E^2}=\frac{2\pi h^3(1+2h^2)J y^5}{(w^2-1)(2h^2-1)\sqrt{B}\,(B+w^2)^2},
    \label{eq:gEE_ads}
    \end{equation}
    \begin{equation}
    G=\frac{\chi_E}{E} =-\frac{y^3 K}{8(B+w^2)J}\,=-\frac{(y-y_{\rm ISCO})(y+y_{\rm ISCO})}{(B+w^2)J},
    \label{eq:gww_ads}
\end{equation}
we have
\begin{equation}
    ds_{\rm orb}^2 =F\,dE^2+ G\,dw^2=\frac{C_w}{E^2}\,dE^2-\frac1E\left(\frac{\partial\lambda}{\partial w}\right)_E\,d w^2.
    \label{eq:Orb metric 5D}
\end{equation}
The two response functions therefore control the two sectors of the orbital geometry. The angular component vanishes at $K=0$, whereas the energy component becomes singular at the minimum temperature locus. The curvature is then obtained from the universal expression eqn. \eqref{eq:curvature_general} in terms of $(y,w)$. Its singularity structure can be understood without its full closed form, and is confirmed by the figures below. To understand the physically significant regions of the curvature, we first define the product of the two metric components as in eqn.~(\ref{eq:FG_product}),
\begin{equation}
    FG = \frac{C_w\chi_E}{E^3} =-\frac{\pi h^3(1+2h^2) K y^8}{4(w^2-1)(2h^2-1)\sqrt{B}\,(B+w^2)^3}.
    \label{eq:FG_ads}
\end{equation}
We of course restrict to the finite radius timelike region $B>0$, where we also have $J, E>0$. Let us now consider the important regions one by one:
\begin{itemize}
\item \textbf{ISCO $(K=0)$:} 
Since $\chi_E\propto -K$, the angular metric component behaves as $G\propto -K$,
whereas $F$ remains finite. Hence,
\begin{equation}
    \frac{1}{FG}\sim K^{-1},
    \quad {\rm and} \quad
    \partial\ln|FG|\sim K^{-1}.
\end{equation}
Therefore the leading contribution to the orbital curvature scales as
\begin{equation}
    R\sim \frac{1}{K^{2}}\sim\frac{1}{\left(y-y_{\rm ISCO}\right)^2}\sim\frac{1}{\left(E-E_{\rm ISCO}\right)^{2}}\,,
    \label{eq:R5_E}
\end{equation}
where $y_{\rm ISCO}$ is given in eqn.~(\ref{eq:yISCO}). Here we used the fact that, at fixed $w$, both $K$ and the energy difference $E-E_{\rm ISCO}$ are linear in $y-y_{\rm ISCO}$, with $E_{\rm ISCO}=w^3/[(w^2-1)\sqrt{w^2+2}]$ from eqn.~(\ref{eq:orbit quantitites_5D}). Thus, the curvature singularity at the ISCO is controlled by the vanishing of the angular response $\chi_E$ (see eqn.~\eqref{eq:gww_ads}), not by that of $C_\Omega$. This is one of the principal differences from the four dimensional Schwarzschild
case. Moreover, $(E,w)$ remain good coordinates at the ISCO, since $E_{,y}\neq 0$ there (see appendix~\ref{app:curvature-5d}), so the divergence can be read off directly in these variables. In the four dimensional case, where $dE/dx$ vanishes at the ISCO, the same conclusion was reached by approaching the ISCO along each branch in the $(x,\Omega)$ coordinates. 
Since $R$ is a scalar, in both cases its divergence is a genuine geometric singularity. The curvature thus diverges quadratically in the distance (in energy) from the ISCO, with the same sign on both sides of it, with the same scaling exponent of $2$ in both four (eqn. (\ref{eq:sch_R_EISCO})) and five dimensions (eqn. (\ref{eq:R5_E})). 

\item \textbf{Minimum temperature locus:}
At the minimum temperature locus, where $2h^2-1=0$ as in eqn. (\ref{tmin1}),  a second type of curvature singularity occurs. First we see that
\begin{equation}
    C_w\sim \frac{1}{(2h^2-1)},
    \qquad
    F=\frac{C_w}{E^2}\sim  \frac{1}{(2h^2-1)},
\end{equation}
while $G$ remains finite. The derivatives of $F$ in eqn. ~\eqref{eq:curvature_general} in the curvature expression then provide the leading singular contribution. Including the overall factor $1/(FG)$ gives
\begin{equation}
    R\sim \frac{1}{(2h^2-1)^2}.
    \label{eq:R5_Tmin}
\end{equation}
Hence the second curvature singularity coincides with the divergence of the
orbital response $C_\Omega$. In this case the singularity is inherited from
the thermodynamics, i.e., from the minimum Hawking temperature point of the background black hole. As at the ISCO, $R$ diverges with the same sign on both sides of this locus.

\item \textbf{Timelike boundary $(B\rightarrow0)$:} 
Here the response behaves as $C_w\sim B^{-3/2}$, but at the same time $E\sim B^{-1/2}$. Thus,
\begin{equation}
    F=\frac{C_w}{E^2}\sim B^{-1/2},
\end{equation}
while $G$ remains finite. The boundary $B=0$ itself is the photon sphere, and lies outside the physical region $B>0$, and the circular timelike orbits only approach it. As it is approached, the leading powers of $B$ cancel in the curvature, and $R$ tends to a finite value. Thus, the divergence of the
response function as the timelike boundary is approached, does not produce an additional curvature
singularity.

\item \textbf{The small $y$ limit:} For any fixed $w>1$, the condition $B=y^2+2-2w^2>0$ requires $y^2>2(w^2-1)$. Hence, the limit $y\to0$ lies outside the physical timelike circular-orbit region. Consequently, the response functions $C_w$ and $\chi_E$ do not describe physical circular orbits in this limit, and any formal divergence of the orbital curvature there should not be interpreted as a physical curvature singularity.
\end{itemize}
The features deduced above can also be confirmed from the exact expression for the curvature. As the expression is quite long and otherwise not very illuminating, the different limiting behaviors outlined above are illustrated in figures~\ref{fig:R_5D}--\ref{fig:Comega_ads}.  Details of the way to obtain the closed form of the expression are relegated to appendix~\ref{app:curvature-5d}.
\noindent
We should mention that, in figure~\ref{fig:R_5D}, for the value of $w$ chosen, $T_{\rm min}$ occurs on the saddle branch. For other values of $w$, $T_{\rm min}$ may instead occur on the center branch, or coincide with the ISCO. This can be understood as follows. As seen in eqn.~(\ref{tmin1}), the minimum temperature locus occurs at $h^2=\frac12$, and from eqn.~(\ref{constraint}) it is located in the orbital variables at
\[
y_{T_{\min}}^4=\frac{4(w^2-1)}{3}.
\]
Since the orbital geometry is defined on physical circular orbits, this locus is relevant when it lies in the timelike region $B>0$, i.e., when $y_{T_{\min}}^4>4(w^2-1)^2$. This holds for
\begin{equation}
1<w<w_c\,, \qquad w_c^2=\frac{4}{3}\,,
\label{eq:wc}
\end{equation}
where $w_c$ is the limiting angular velocity of circular orbits approaching the photon sphere of the minimum temperature black hole. Indeed, around this black hole circular timelike orbits exist only outside its photon sphere, and their angular velocities fill precisely the range $1<w<w_c$. For $w\geq w_c$, the fixed-$w$ family consists entirely of small black holes, $h<1/\sqrt{2}$, and the ISCO is then the only finite radius curvature singularity. For $1<w<w_c$, whether the minimum temperature point belongs to the center or saddle branch is determined independently by the orbital stability factor $K$ given in eqn.~(\ref{K_5D}), which at the minimum temperature locus is
\[
K_{T_{\min}}
=
\frac{8}{y_{T_{\min}}^3}\left(
2+\sqrt{\frac{4(w^2-1)}{3}}-w^2-w^4\right).
\]
The two cases are separated by $w=w_*$, where $K_{T_{\min}}=0$, or equivalently
\[
(w_*^2-1)(w_*^2+2)^2=\frac43\,.
\]
This is the angular velocity at the ISCO of the minimum temperature black hole. Hence
\[
\begin{aligned}
&1<w<w_*: && T_{\min}\ \text{lies on the center branch},\\
&w_*<w<w_c: && T_{\min}\ \text{lies on the saddle branch}.
\end{aligned}
\]
Numerically, $w_c\simeq 1.155$ and $w_*\simeq 1.066$, so that the value $w=1.1$ used in figure~\ref{fig:R_5D} corresponds to the saddle branch case. Not surprisingly then, the curvature singularity at $T_{\min}$ is fundamentally a background thermal response singularity rather than a center/saddle feature. At $w=w_*$, the minimum temperature and ISCO singularities coincide. Nevertheless, in any of these situations, the fact remains that the curvature diverges, both at $T_{\min}$ and at the ISCO.

\noindent
{\bf On response functions:}
It is useful to emphasize that the two response functions appearing in
the orbital metric and given in eqns. (\ref{eq:Cw_ads}) and (\ref{ChiE_5D}) probe different aspects of the system, which is also seen from their contrasting behaviors in figures~\ref{fig:chiE_ads} and~\ref{fig:Comega_ads}, respectively. The behavior of angular susceptibility in figure~\ref{fig:chiE_ads} can be explained as follows. From eqn. (\ref{ChiE_5D}),   in the physical timelike
region, we have $y>0$, $B>0$ and $J>0$, and thus 
$
 \operatorname{sgn}(\chi_E)=-\operatorname{sgn}(K)
$, in addition to $\chi_E$ vanishing precisely at the ISCO, as displayed in
figure~\ref{fig:chiE_ads}.  Its sign therefore directly tracks the
center-saddle character of the circular orbit. The response $C_\Omega$, on the other hand, probes the variation of
the orbital energy with the Hawking temperature of the background at
fixed angular velocity. Hence, from eqn. (\ref{eq:Cw_ads}), 
$
 \operatorname{sgn}(C_\Omega)
 =
 \operatorname{sgn}(2h^2-1).
$
Therefore,
\begin{equation}
 C_\Omega>0
 \qquad\text{for}\qquad
 h>\frac{1}{\sqrt{2}},
\end{equation}
corresponding to the large black hole branch, while
\begin{equation}
 C_\Omega<0
 \qquad\text{for}\qquad
 h<\frac{1}{\sqrt{2}},
\end{equation}
corresponding to the small black hole branch. 
At
$h=1/\sqrt{2}$, the Hawking temperature reaches its minimum and
$C_\Omega$ diverges, as shown in figure~\ref{fig:Comega_ads}. The two response functions therefore diagnose two independent
properties. While $\chi_E$ keeps track of orbital stability, $C_\Omega$ characterizes the black hole branches. The different possible combinations of signs of response functions are collected in table~\ref{table1}. We see that unlike thermodynamic susceptibilities, whose signs are fixed by convexity of a potential, the contrasting signs of $C_\Omega$ and $\chi_E$ for black hole branches and orbits are a feature of the orbital geometry. In particular, $\chi_E$ is evaluated at fixed orbital energy and therefore mixes orbital and background variations. 
\noindent
\begin{table}[htbp]
\centering
\caption{Signs of the two orbital
response functions based on orbital and black hole branch. Large black holes ($h>1/\sqrt{2}$) occur in the timelike region for $1<w<w_c$, and on the center branch for $1<w<w_*$.}
\label{tab:response-signs}
\renewcommand{\arraystretch}{1.35}
\setlength{\tabcolsep}{12pt}
\begin{tabular}{|c|c|c|}
\hline
\textbf{Orbital branch} &
\textbf{Black hole branch} &
$\boldsymbol{(\chi_E,C_\Omega)}$\\
\hline
Center: $K>0$ &
Large: $h>1/\sqrt{2}$ &
$(-,+)$\\
\hline
Center: $K>0$ &
Small: $h<1/\sqrt{2}$ &
$(-,-)$\\
\hline
Saddle: $K<0$ &
Large: $h>1/\sqrt{2}$ &
$(+,+)$\\
\hline
Saddle: $K<0$ &
Small: $h<1/\sqrt{2}$ &
$(+,-)$\\
\hline
\end{tabular}
\label{table1}
\end{table}


\noindent
To summarize, the five dimensional Schwarzschild AdS example therefore exhibits two distinct mechanisms in the orbital curvature. One of them is associated with an orbital marginal stability condition, i.e. the ISCO singularity. This is characterized by the vanishing of $\chi_E$ and the resulting curvature divergence as $R\sim K^{-2}$, while the second one, present for $1<w<w_c$, is associated with the thermodynamic structure of the Schwarzschild AdS background. That is,  the minimum temperature locus is characterized by the divergence of $C_\Omega$ and $R\sim(2h^2-1)^{-2}$. No further finite radius curvature singularities arise from the factors $B$, $J$ or $E$ within the physical timelike region. The orbital geometry thus captures both the marginal stability of the particle orbit and the thermodynamic structure of the AdS black hole background.

\section{Conclusion}
\label{sec:conclusion}

In this work, we proposed an orbital response geometry for circular timelike orbits around general black holes, with the specific intention of probing the ISCOs. Using the fact that the circular orbits are characterized  by a few quantities such as energy, angular momentum and angular velocity, with the latter two being conjugate to each other, we started with an ansatz for the orbital metric in eqn. (\ref{eq:orbital_metric}), which contained a scaling factor $\alpha$ with dimensions of inverse length. The metric contains two orbital response functions, namely, $C_{\Omega}$ (through the combination $\alpha C_\Omega$) and $\chi_E$, which characterize, respectively, the sensitivity of the orbital energy to changes in the black hole background and the angular momentum response to changes in orbital angular velocity. The corresponding scalar curvature $R$ of the orbital geometry obtained in eqn. (\ref{eq:curvature_general}) is also very general: it provides a geometric diagnostic of bifurcation points of the circular orbits and also captures the degenerations in the orbital response. We then studied specific examples corresponding to the four and five dimensional Schwarzschild black holes in asymptotically flat and AdS spacetimes. For the asymptotically flat Schwarzschild black hole, the only available scale is the black hole mass, and we thus take $\alpha=T_H$, whereas for the AdS case, the natural length scale allowed us to pick $\alpha = 1/\ell$.  In both examples, we found that the ISCO is singled out by a curvature singularity, although through different response channels. In the AdS case an additional curvature singularity is associated with the minimum temperature point of the black hole background, and is therefore distinct from the orbital ISCO singularity. Thus, the curvature does not merely reproduce the response functions: it selects which of their singularities are geometric, it encodes their scaling through invariant exponents, and as we noted in particular in five dimensions, it combines the orbital and background sectors into a single invariant.\\

\noindent
The sign and zero structure of the curvature $R$ further partitions the physically allowed orbital domain. In four dimensions, $R$ vanishes once, on the stable branch (for the choice $\alpha=T_H$), while the ISCO separates the center and saddle sectors. These results show that the orbital geometry may be viewed as a geometric framework for organizing orbital response, with the ISCO appearing as a distinguished singular locus of this geometry.  The orbital response functions $C_{\Omega}$ and $\chi_E$ which appeared, respectively, in eqns. (\ref{eq:COmega_general}) and (\ref{eq:chiE_general}) during the construction of the orbital geometry, also capture the ISCO.  For the four dimensional Schwarzschild case, $C_{\Omega}$ changes sign between the center and saddle branches and vanishes at the ISCO,
\begin{equation}
C_{\Omega}>0\quad{\rm(center)},\qquad
C_{\Omega}=0\quad{\rm(ISCO)},\qquad
C_{\Omega}<0\quad{\rm(saddle)}.
\end{equation}
The curvature singularity is therefore associated with the degeneration of the energy response sector. We also noted that  the first quantity, $C_{\Omega}$, measures the response of the energy of the circular orbit to a variation of the black hole background, parametrized by its Hawking temperature, while keeping the orbital angular velocity fixed.  Thus it is distinct from the usual heat capacity of the black hole, which is generally defined as
\begin{equation}
C_{\rm BH}=\frac{dM}{dT_H}.
\end{equation}
Physically, changing $T_H$ changes the mass or horizon scale of the background, and hence the gravitational environment experienced by the orbit. Thus, $C_{\Omega}$ measures the corresponding change in the orbital energy required to maintain fixed $\Omega$. It would be interesting to see if these features are generic by studying more examples.\\

\noindent
The second response function $\chi_E$
may be viewed as an angular susceptibility.  It measures how the angular momentum of a circular orbit responds to a change in angular velocity at fixed orbital energy.  This quantity has a particularly direct connection with orbital stability in the five dimensional Schwarzschild AdS example, where we found that $\chi_E\propto -K$,
with $K$ denoting the marginal stability factor.  As a result, the sign of $\chi_E$ captures the stability structure of circular orbits, i.e.,
\begin{equation}
\chi_E<0\quad{\rm(center/stable)},\qquad
\chi_E=0\quad{\rm(ISCO)},\qquad
\chi_E>0\quad{\rm(saddle/unstable)}.
\end{equation}
We also found that in the five dimensional Schwarzschild AdS example, $C_{\Omega}$ remains finite at the ISCO, whereas $\chi_E$ vanishes there.  The ISCO is therefore detected through the angular response sector.  A separate divergence of $C_{\Omega}$ occurs at the minimum temperature point of the Schwarzschild AdS background, and represents a background thermodynamic effect rather than the orbital instability itself. \\

\noindent
We end this manuscript with some reflections on how the orbital geometry developed here relates to, and differs from, the Ruppeiner geometry of equilibrium thermodynamics, which partly motivated its construction. The Ruppeiner metric is defined as the Hessian of the entropy with respect to the extensive variables~\cite{ruppeiner1979thermodynamics,ruppeiner1995riemannian}. Its origin lies in equilibrium fluctuation theory: the metric governs the Gaussian probability of fluctuations about an equilibrium state, and is therefore positive definite wherever the system is locally stable. This origin gives its scalar curvature a direct physical meaning. Its magnitude is related to the correlation volume, so that it diverges at critical points with characteristic exponents, while its sign has been argued to reflect whether the effective microscopic interactions are attractive or repulsive, with a flat geometry corresponding to an ideal gas~\cite{Ruppeiner:2010dzw}. In black hole physics, these ideas have led to a rich picture of black hole microstructures~\cite{Wei:2015iwa,Wei:2019uqg}. The thermodynamic interpretation of the curvature is thus anchored in a well understood statistical framework.\\

\noindent
The orbital geometry introduced in the present work shares the spirit of this construction, but rests on a different foundation. Its state space is the family of circular timelike geodesics, rather than an ensemble of equilibrium states, and its metric is built directly from two orbital response functions, $\alpha C_\Omega$ and $\chi_E$, in close analogy with the way heat capacities and compressibilities appear in the components of the Ruppeiner metric. Several features familiar from thermodynamic geometry nevertheless reappear naturally. The degeneration of a response function leads to a curvature singularity (such as at the ISCO), while a divergence in the response function does not necessarily do so (as seen when approaching the photon orbit). Moreover, in four dimensions the signature of the metric tracks orbital stability, much as positive definiteness of the Ruppeiner metric reflects thermodynamic stability. In the AdS case, the orbital curvature further separates singularities associated with marginal orbital stability from those inherited from the thermodynamics of the background, a distinction that does not arise in the purely Ruppeiner geometry. 
In both the four and five dimensional examples studied here, at fixed angular velocity the curvature scales as
$R\sim|E-E_{\rm ISCO}|^{-2}$ near the ISCO (see eqns.~(\ref{eq:sch_R_EISCO})
and~(\ref{eq:R5_E})). This coincides with the mean field form
$R\sim|t|^{-(2-\alpha)}$, $\alpha=0$, of thermodynamic geometry~\cite{ruppeiner1995riemannian,Janke:2004ef}, with $E$
playing the role of temperature $t$, and is consistent with the mean field
exponents of ISCOs found in~\cite{bhamidipati2026criticality}. Whether this
reflects an underlying hyperscaling relation for the orbital geometry
remains an open question. The orbital curvature is therefore already found to be a useful geometric diagnostic of the circular orbit family, though the possible connection to any statistical interpretation, and the generality of its applicability for other black holes, need further investigation.\\

\noindent
Looking forward, the aforementioned comparisons point to several interesting research directions, some of which we mention now. First, it would be valuable to identify a potential whose Hessian reproduces the orbital metric, or a close relative of it. A natural starting point is the relation $dE=\Omega\,dL$ obeyed along circular orbits, which, once supplemented by a variation of the background, assigns to the orbital energy the role of a fundamental relation with $(\Omega,L)$ as a conjugate pair. Second, the orbital metric proposed in eqn. (\ref{eq:orbital_metric}) was motivated by an analogy with thermodynamic geometry. A first principles derivation of the metric, for instance from a suitable ensemble or fluctuation analysis of nearly circular geodesics, could supply the orbital counterpart of the fluctuation theoretic origin of the Ruppeiner metric. It may also connect the orbital curvature to dynamical quantities such as epicyclic frequencies and Lyapunov exponents of circular orbits~\cite{Cardoso:2008bp}. Third, it would be interesting to understand whether the sign and magnitude of the orbital curvature admit a physical interpretation analogous to the interaction and correlation volume pictures of thermodynamic geometry. Finally, as in thermodynamic geometry~\cite{Quevedo:2006xk,Bravetti:2012dn}, the dependence of the construction on the choice of coordinates and response functions deserves a systematic study, for which Legendre invariant formulations may provide useful guidance. \\

\noindent
It would of course be interesting to extend our construction to more general situations involving charged black holes in Einstein gravity and higher derivative theories, where interesting phase transitions occur~\cite{Chamblin:1999tk,Kubiznak:2012wp,Kubiznak:2016qmn}. In such cases, it is a genuine question whether the orbital geometry constructed here captures not just the stability features of circular orbits, but can also be generalized to detect the critical points of black holes.  Extending the construction to higher dimensions would put it to more stringent tests. For instance, for the asymptotically flat Schwarzschild black holes in spacetime dimension greater than four, it is well known that  there are no stable circular orbits or ISCOs~\cite{Tangherlini:1963bw,Cardoso:2008bp}. Thus, the curvature is not expected to have any singularities.  However, for higher dimensional AdS Schwarzschild black holes, one expects both stable circular orbits and ISCOs, and hence, the corresponding curvature is expected to diverge at these special points. It is thus worth checking these features. In this direction, one of the better tests of our construction would also be to extend it to non-static, rotating black holes, such as the Kerr and Kerr-AdS black holes, where the equatorial circular orbits split into co-rotating and counter-rotating families with distinct ISCOs~\cite{Bardeen1972}, and the background acquires an additional thermodynamic parameter, the angular momentum. It would be equally interesting to consider genuinely non-stationary, dynamical spacetimes, such as accreting Vaidya type black holes, in which the ISCO itself evolves in time~\cite{song2021evolutions}.

\section*{Acknowledgements}
CB thanks ANRF for the ARG-MATRICS grant no. ANRF/ARGM/2025/002280/MTR. CB thanks Renann Lipinski Jusinskas for hospitality at FZU and CEICO, Prague, and the European Structural
and Investment Funds, and the Czech Ministry of Education, Youth and Sports
(project FORTE CZ.02.01.01/00/22-008/0004632) for partial support.

\appendix
\section{Derivation of the orbital curvature}
\label{app:curvature}

We consider circular timelike orbits in a one parameter family of static,
spherically symmetric black holes with the metric of the form shown in eqn. (\ref{eq:metric}) and reproduced here for clarity:
\begin{equation}
    ds^2
    =-f(r;p)\,dt^2 + \frac{dr^2}{f(r;p)} + r^2d\Sigma_{d-2}^2.
\label{eq:metric1}
\tag{A1}
\end{equation}
Let \(p\) denote the parameter
specifying the background, and let \(r_c\) denote the radius of a circular
orbit.  The corresponding orbital energy, angular momentum and angular
velocity may therefore be regarded as functions
\begin{equation}
E=E(r_c,p),\qquad
L=L(r_c,p),\qquad
\Omega=\Omega(r_c,p),
\tag{A2}
\end{equation}
while the Hawking temperature of the background is denoted by \(T_H=T_H(p)\).
The pair \((r_c,p)\) thus provides a convenient parametrization of the
two-dimensional family of circular orbits. In proposing our orbital metric in eqn. (\ref{eq:orbital_metric}), we actually used the orbital variables \((E,\Omega)\) as
coordinates on this state space.  The corresponding orbital line element was
taken to be
\begin{equation}
ds^2
=
F(E,\Omega)\,dE^2
+
G(E,\Omega)\,d\Omega^2,
\qquad
F=\frac{\alpha\,C_\Omega}{E^2},
\qquad
G=\frac{\chi_E}{E},
\tag{A3}
\end{equation}
where
\begin{equation}
C_\Omega
=
\left(\frac{\partial E}{\partial T_H}\right)_\Omega,
\qquad
\chi_E
=
-\left(\frac{\partial L}{\partial\Omega}\right)_E .
\tag{A4}
\end{equation}
For practical calculations it is not necessary to solve explicitly for
\(r_c\) and \(p\) as functions of \(E\) and \(\Omega\). We work around this by introducing a
Jacobian as
\begin{equation}
\mathcal{J}[P,Q]
=
\frac{\partial(P,Q)}{\partial(r_c,p)}
=
P_{,r_c}Q_{,p}-P_{,p}Q_{,r_c}.
\tag{A5}
\end{equation}
Using the standard Jacobian identities, the response functions can then be written
directly in terms of derivatives with respect to \((r_c,p)\) as:
\begin{equation}
C_\Omega
=
\frac{\mathcal{J}[E,\Omega]}
     {\mathcal{J}[T_H,\Omega]},
\qquad
\chi_E
=
-\frac{\mathcal{J}[E,L]}
       {\mathcal{J}[E,\Omega]}.
\tag{A6}
\end{equation}
Here, we can think of \(T_H\) as a function on the \((r_c,p)\) space with no
explicit dependence on \(r_c\). The same Jacobian transformation also allows derivatives with respect to the
orbital coordinates \((E,\Omega)\) to be evaluated parametrically.  Although the use of Jacobians is not essential for the simpler four dimensional computation discussed in section~\ref{sec:schwarzschild}, it considerably simplifies the corresponding calculations in the  five dimensional example in section~\ref{sec:schwarzschild_ads}. Further, for any
function of \((r_c,p)\),
\begin{equation}
\left.\frac{\partial}{\partial E}\right|_\Omega
=
\frac{
\Omega_{,p}\,\partial_{r_c}
-
\Omega_{,r_c}\,\partial_p
}{
\mathcal{J}[E,\Omega]
},
\qquad
\left.\frac{\partial}{\partial\Omega}\right|_E
=
\frac{
-E_{,p}\,\partial_{r_c}
+
E_{,r_c}\,\partial_p
}{
\mathcal{J}[E,\Omega]
}.
\tag{A7}
\end{equation}
Equations (A6) and (A7) are useful because all quantities entering the
orbital curvature can also now be calculated in the original variables
\((r_c,p)\), without performing an explicit inversion to \((E,\Omega)\). We now derive the scalar curvature of the metric (A3).  Since the metric is
diagonal,
\begin{equation}
g_{EE}=F,\qquad
g_{\Omega\Omega}=G,
\qquad
g^{EE}=\frac{1}{F},\qquad
g^{\Omega\Omega}=\frac{1}{G}.
\tag{A8}
\end{equation}
The non-vanishing Christoffel symbols are
\begin{align}
\Gamma^{E}_{EE}
&=\frac{F_{,E}}{2F},
&
\Gamma^{E}_{E\Omega}
&=\frac{F_{,\Omega}}{2F},
&
\Gamma^{E}_{\Omega\Omega}
&=-\frac{G_{,E}}{2F},
\tag{A9}
\\
\Gamma^{\Omega}_{EE}
&=-\frac{F_{,\Omega}}{2G},
&
\Gamma^{\Omega}_{E\Omega}
&=\frac{G_{,E}}{2G},
&
\Gamma^{\Omega}_{\Omega\Omega}
&=\frac{G_{,\Omega}}{2G}.
\tag{A10}
\end{align}
Here a comma denotes differentiation with respect to the indicated
state space coordinate, like
\(F_{,E}=(\partial F/\partial E)_\Omega\) and so on. Substitution into the definition of the Riemann tensor gives the single
independent component
\begin{align}
R_{E\Omega E\Omega}
={}&
-\frac{1}{2}
\left(
G_{,EE}+F_{,\Omega\Omega}
\right)
+
\frac{1}{4}G_{,E}
\left(
\frac{F_{,E}}{F}
+
\frac{G_{,E}}{G}
\right)
+
\frac{1}{4}F_{,\Omega}
\left(
\frac{F_{,\Omega}}{F}
+
\frac{G_{,\Omega}}{G}
\right).
\tag{A11}
\end{align}
For a two dimensional metric the scalar curvature is related to this
component by $ R = 2R_{E\Omega E\Omega}/FG$, and therefore
\begin{equation}
R
=
-\frac{1}{FG}
\left[
G_{,EE}
+
F_{,\Omega\Omega}
-
\frac{1}{2}G_{,E}\,
\partial_E\ln|FG|
-
\frac{1}{2}F_{,\Omega}\,
\partial_\Omega\ln|FG|
\right].
\tag{A12}
\end{equation}
This is the general expression used in the main text in eqn. (\ref{eq:curvature_general}). Equations (A6), (A7) and (A12) provide a complete
parametric prescription for computing the orbital curvature.  Starting from
the quantities characterizing the circular orbit family, such as, 
\(E(r_c,p)\), \(L(r_c,p)\), \(\Omega(r_c,p)\) and the background temperature
\(T_H(p)\), the response functions and all derivatives entering \(R\) can be
evaluated directly in the \((r_c,p)\) parametrization. This was done to obtain an analytical expression for the curvature, particularly in situations where the solutions
for \(r_c(E,\Omega)\) or \(p(E,\Omega)\), for general black holes are difficult to obtain.

\subsection{Schwarzschild black holes in four dimensions}
\label{app:curvature-4d}

\setcounter{equation}{0}
\renewcommand{\theequation}{A1.\arabic{equation}}
\renewcommand{\theHequation}{A1.\arabic{equation}}

We use the metric functions $F=g_{EE}$ and $G=g_{\Omega\Omega}$ given in
eqns.~\eqref{eq:sch_gEE} and~\eqref{eq:sch_gOmegaOmega}, together with
eqn.~\eqref{eq:sch_dEdx}.  For a diagonal metric, it is convenient to express the general form of the scalar curvature given in eqn. (\ref{eq:curvature_general}) for the present case as
\begin{equation}
R=R_1+R_2+R_3+R_4+R_5+R_6\, ,
\end{equation}
with
\begin{equation}
\begin{aligned}
R_1={}&-\frac{G_{,EE}}{FG},
\qquad
R_2=-\frac{F_{,\Omega\Omega}}{FG},
\qquad
R_3=\frac{F_{,\Omega}G_{,\Omega}}{2FG^2},\\
R_4={}&\frac{G_{,E}^2}{2FG^2},
\qquad
R_5=\frac{F_{,E}G_{,E}}{2F^2G},
\qquad
R_6=\frac{F_{,\Omega}^2}{2F^2G}.
\end{aligned}
\label{eq:app-4d-sixpieces}
\end{equation}
Since $E=E(x)$, differentiation at fixed $\Omega$ is obtained from
\[
\left(\frac{\partial}{\partial E}\right)_\Omega
=\frac{1}{E_{,x}}\partial_x
=\frac{2(1-3x)^{3/2}}{6x-1}\,\partial_x ,
\]
while at fixed $E$ one simply has
\[
\left(\frac{\partial}{\partial\Omega}\right)_E=\partial_\Omega .
\]
All six parts of the curvature in eqn.~\eqref{eq:app-4d-sixpieces} can now be evaluated as follows. For
$R_1, R_4$, the required derivatives are
\[
G_{,E}=\frac{G_{,x}}{E_{,x}}
=
\frac{2(1-3x)^{3/2}}
{\Omega^2(1-2x)^2(6x-1)},
\]
and
\[
G_{,EE}
=
\frac{G_{,xx}}{E_{,x}^2}
-\frac{G_{,x}E_{,xx}}{E_{,x}^3},
\qquad
E_{,xx}=\frac{3(6x+1)}{4(1-3x)^{5/2}} .
\]
For $R_2, R_3, R_6$ terms, the required explicit $\Omega$ derivatives are,
\[
F_{,\Omega}=0,
\qquad
F_{,\Omega\Omega}=0,
\qquad
G_{,\Omega}=-\frac{2G}{\Omega},
\]
since $F$ depends only on $x$, which is held fixed together with $E$. Hence the three terms involving $F_{,\Omega}$ vanish identically,
\begin{equation}
R_2=R_3=R_6=0 ,
\label{eq:app-4d-R236}
\end{equation}
and the curvature reduces to $R=R_1+R_4+R_5$.
Finally, the derivative entering $R_5$ is
\[
F_{,E}=\frac{F_{,x}}{E_{,x}},
\qquad
\frac{F_{,x}}{F}
=
\frac{36x^3+36x^2-23x+2}
{2x(1-6x)(1-2x)(1-3x)} .
\]
Substitution into eqn.~\eqref{eq:app-4d-sixpieces} gives the remaining
contributions
\begin{equation}
\label{eq:app-4d-R1}
R_1=
-\frac{6(1-3x)^{5/2}
\left(108x^2-60x+11\right)}
{x^{2}(6x-1)^4}\, ,
\qquad
R_4=
-\frac{6(1-3x)^{7/2}}
{x^{3}(6x-1)^3}\, ,
\end{equation}
\begin{equation}
R_5=
\frac{3(1-3x)^{5/2}
\left(36x^3+36x^2-23x+2\right)}
{x^{3}(6x-1)^4}\, .
\label{eq:app-4d-R5}
\end{equation}
Adding eqns.~(\ref{eq:app-4d-R1}) and (\ref{eq:app-4d-R5}), $R=R_1+R_4+R_5$ reproduces the closed expression for $R$ given in eqn.~\eqref{eq:sch_R}.

\subsection{AdS Schwarzschild black holes in five dimensions}
\label{app:curvature-5d}

\setcounter{equation}{0}
\renewcommand{\theequation}{A2.\arabic{equation}}
\renewcommand{\theHequation}{A2.\arabic{equation}}

For the present case, we use directly the metric functions $F$ and
$G$ given in eqns.~\eqref{eq:gEE_ads} and~\eqref{eq:gww_ads}.
Differentiating the orbital energy gives
\[
E_{,y}=-\frac{2J}{y^2B^{3/2}},
\qquad
E_{,w}=\frac{2w^3}{yB^{3/2}},
\]
and therefore
\begin{equation}
{\cal D}_E\equiv
\left(\frac{\partial}{\partial E}\right)_w
=
-\frac{y^2B^{3/2}}{2J}\,\partial_y,
\qquad
{\cal D}_w\equiv
\left(\frac{\partial}{\partial w}\right)_E
=
\partial_w+\frac{w^3y}{J}\,\partial_y .
\label{eq:app-5d-operators}
\end{equation}
The implicit derivatives of $h(y,w)$ obtained using
eqn.~\eqref{constraint} are
\begin{equation}
h_{,y}
=-\frac{2h(1+h^2)}{y(1+2h^2)},
\qquad
h_{,w}
=\frac{w}{h(1+2h^2)y^4}.
\label{eq:app-5d-hderivatives}
\end{equation}
In eqn.~\eqref{eq:app-5d-operators}, $\partial_y$ and $\partial_w$ also act on the implicit dependence on $h(y,w)$ through eqn.~\eqref{eq:app-5d-hderivatives}. Unlike the four dimensional case, $F$ now depends on $w$ both explicitly and through $h$, so none of the terms vanish identically. Using these derivative operators, the five dimensional scalar curvature
can once again be written as the sum of six terms,
\begin{equation} \label{R5}
R= R_1+R_2+R_3+R_4+R_5+R_6,
\end{equation}
where
\begin{equation}
\begin{aligned}
R_1&=-\frac{1}{F}
\left({\cal D}_EQ_E+Q_E^2\right),
&
R_2&=-\frac{1}{G}
\left({\cal D}_wP_w+P_w^2\right),
\\[1mm]
R_3&=\frac{P_wQ_w}{2G},
&
R_4&=\frac{Q_E^2}{2F},
\\[1mm]
R_5&=\frac{P_EQ_E}{2F},
&
R_6&=\frac{P_w^2}{2G}\, .
\end{aligned}
\label{eq:app-5d-sixpieces}
\end{equation}
The quantities entering eqn.~\eqref{eq:app-5d-sixpieces} are the
logarithmic derivatives of the two metric functions,
\begin{equation} \label{PQ}
P_E={\cal D}_E\ln |F|,\qquad
Q_E={\cal D}_E\ln |G|,\qquad
P_w={\cal D}_w\ln |F|,\qquad
Q_w={\cal D}_w\ln |G| .
\end{equation}
We first emphasize that $R_1$ and $R_2$ in eqn.~\eqref{eq:app-5d-sixpieces} contain the
second derivatives of the metric functions.  Since
\[
G_{,E}=GQ_E,
\qquad
F_{,w}=FP_w,
\]
one obtains
\[
G_{,EE}
=
G\left({\cal D}_EQ_E+Q_E^2\right),
\qquad
F_{,ww}
=
F\left({\cal D}_wP_w+P_w^2\right),
\]
which explains directly the forms of $R_1$ and $R_2$ in
eqn.~\eqref{eq:app-5d-sixpieces}.  The remaining four terms, i.e.,  $R_3, R_4, R_5$ and $R_6$  follow
from the first derivatives
\[
F_{,E}=FP_E,\qquad
G_{,E}=GQ_E,\qquad
F_{,w}=FP_w,\qquad
G_{,w}=GQ_w\, .
\]
Finally, the required logarithmic derivatives in eqn. (\ref{PQ}) are 
\[
P_E=
-\frac{y^2B^{3/2}}{2J}
\left[
\frac{1+2\Theta}{y}
+\frac{2y}{J}
-\frac{y}{B}
-\frac{4y}{B+w^2}
\right],
\]
\[
Q_E=
\frac{y^3B^{3/2}}{J}
\left[
-\frac{8}{y^3K}+\frac1J+\frac1{B+w^2}
\right],
\]
\[
\begin{aligned}
P_w={}&
-\frac{w\Theta}{w^2-1}
+\frac{2w(2w^2-3)}{J}
+\frac{2w}{B}
+\frac{4w}{B+w^2}
\\
&+\frac{w^3y}{J}
\left[
\frac{1+2\Theta}{y}
+\frac{2y}{J}
-\frac{y}{B}
-\frac{4y}{B+w^2}
\right],
\end{aligned}
\]
and
\[
\begin{aligned}
Q_w={}&
-\frac{16w(2w^2+1)}{y^3K}
-\frac{2w(2w^2-3)}{J}
+\frac{2w}{B+w^2}
\\
&-\frac{2w^3y^2}{J}
\left[
-\frac{8}{y^3K}+\frac1J+\frac1{B+w^2}
\right]\, ,
\end{aligned}
\]
where we defined
\[
\Theta(h)=
\frac{4h^6+4h^4+7h^2+1}
{(2h^2-1)(1+2h^2)^2}.
\]
These expressions follow directly from eqns.~\eqref{eq:app-5d-operators}
and~\eqref{eq:app-5d-hderivatives}.  Substitution into
eqn.~\eqref{eq:app-5d-sixpieces} gives all the required terms of the
curvature in eqn.~\eqref{R5}, which were used for generating the figures in section~\ref{sec:schwarzschild_ads}.
\addcontentsline{toc}{section}{References}
\bibliographystyle{JHEP}

\bibliography{bib}

\end{document}